%% file: PRL-Dhh_v0.8.tex
\documentclass[aps, prl,twocolumn,superscriptaddress,showpcs,amsmath,amssymb]{revtex4} 

\usepackage{mathrsfs}
\usepackage{graphicx,overpic,rotating}
\usepackage{hyperref}
\usepackage{dcolumn}
\usepackage{bm}
\usepackage{xspace}
\usepackage{multirow}
\usepackage{booktabs}
\usepackage{longtable}
\usepackage{lineno}

\input{macro}

\begin{document}
\graphicspath{{plots_for_paper/}}
\title{Measurement of  the difference of CP--violating asymmetries in $D^0\to K^+K^-$ and $D^0\to \pi^+\pi^-$ decays at CDF}
\input{September2011_Authors}
\date{\today}

\begin{abstract}
We report a measurement of the difference ($\Delta\Acp$) between time-integrated \CP--violating asymmetries  in \mbox{$D^0\to K^+ K^-$} and \mbox{$D^0\to \pi^+\pi^-$} decays reconstructed in the full data set  of proton-antiproton collisions collected by the Collider Detector at Fermilab, corresponding to  $9.7$~fb$^{-1}$ of integrated luminosity. The strong decay $D^{*+}\to D^0\pi^+$  is used to identify the charm meson at production as $D^0$ or $\overline{D}^0$. 
 We measure $\Delta\Acp = \bigl(-0.62 \pm 0.21 \stat \pm 0.10 \syst\bigr)\%$, which differs from zero by 2.7 Gaussian standard deviations. 
 This result supports similar evidence for \CP\ violation in charm-quark decays obtained in proton-proton collisions.
\end{abstract}


\maketitle

The noninvariance of the laws of physics under the simultaneous transformations of parity and charge conjugation (\CP\ violation) is accommodated in the standard model (SM) through the presence of a single irreducible complex phase in the weak couplings of quarks. Generic extensions of the SM are expected to introduce additional sources of \CP\ violation, which, if observed, could provide indirect signs of new particles or interactions.
 To date, \CP\ violation has been established in transitions of strange and bottom hadrons, with effects consistent with the SM interpretation~\cite{pdg,hfag,Antonelli:2009ws}.
Studies of \CP\ violation in charm decays provide a unique probe for non-SM physics. The neutral $D$ system is complementary to the $B$ and $K$ sectors in that up-type quarks (electric charge +2/3) are involved in the initial state. Therefore, \CP-violating effects probe the presence of down-type (charge $-1/3$) new physics through charged-current couplings~\cite{Bianco:2003vb,Artuso:2008vf,Shipsey:2006zz,Burdman:2003rs}.  However, \CP-violating effects are expected not to exceed $\mathcal{O}(10^{-2})$ in the SM~\cite{Bianco:2003vb}, because charm transitions are well described by the physics of the first two quark generations. Indeed, no \CP--violating effects have been firmly established yet in charm dynamics.\par  
 Time-integrated \CP--violating asymmetries of decays  into \CP\ eigenstates such as $D^0 \to \pi^+\pi^-$ and $D^0 \to K^+ K^-$ probe non-SM physics contributions in the {\it oscillation}  and {\it penguin} transition amplitudes.  Oscillation indicates $D^0$-$\overline{D}^0$ transitions governed by the exchange of virtual heavy particles occurring before the $D$ meson decay. Penguin decays are second-order transitions mediated by an internal loop. Both amplitudes may be affected by the exchange of non-SM particles, which could  enhance the size of the observed \CP\ violation with respect to the SM expectation. 
In 2011, CDF reported \CP-violating asymmetries compatible with zero within a few $10^{-3}$ uncertainty in these decays, along with a measurement of the difference $\Delta\Acp$  of \CP-violating asymmetries  in $D^0\to K^+ K^-$ and $D^0\to\pi^+\pi^-$, also consistent with zero \cite{Aaltonen:2011se}. Shortly after, LHCb reported a more precise determination of the difference, which is $3.5\sigma$ different from zero~\cite{Aaij:2011in}. If established, this result provides the first evidence for \CP\ violation in charm dynamics, with a size larger than most SM expectations~\cite{SM-predictions}. 
 Among the quantities sensitive to \CP\ violation in charm mesons, $\Delta\Acp$ can be measured with good accuracy because many common systematic uncertainties cancel. In addition, $\Delta\Acp$ could be maximally sensitive to \CP\ violation since the individual asymmetries are expected to have opposite signs, if the invariance of the dynamics under interchange of $d$ with $s$ quarks is approximately valid~\cite{Bianco:2003vb}.\par
In this Letter, we report a measurement of the difference of \CP--violating asymmetries  in $D^0\to K^+ K^-$ and $D^0\to \pi^+\pi^-$ decays reconstructed in  the full data set of 1.96 TeV proton-antiproton collisions collected by the Collider Detector at Fermilab, corresponding to  $9.7$~fb$^{-1}$ of integrated luminosity. 
In addition to the increase in the size of the data set with respect to the previous measurement~\cite{Aaltonen:2011se}, we optimize the selection for a measurement of the difference of asymmetries, reaching a sensitivity competitive with the LHCb result~\cite{Aaij:2011in}. \par 
For each decay mode, the \CP--violating time-integrated asymmetry between decays of states produced as $D^0$ and $\Dbar^0$ is defined as
\begin{equation}\label{eq:acp}
\frac{N(D^0\to h^+h^-;t)-N(\Dbar^0\to h^+h^-;t)}{N(D^0\to h^+h^-;t)+N(\Dbar^0\to h^+h^-;t)},
\end{equation}
where $h$ identifies a charged kaon or pion.  
 The asymmetry can receive contributions from any difference in partial decay widths between $D^0$ and $\Dbar^0$ mesons (direct \CP\ violation) and both the difference in mixing probabilities between $D^0$ and $\overline{D}^0$ mesons and the interference between mixed and unmixed decays (indirect \CP\ violation). Due to the slow mixing rate of charm mesons, the asymmetry is approximated to first order as the sum of two terms \cite{Aaltonen:2011se}, \begin{align}\label{eq:acp3}
\Acp(h^+h^-) 	
			&=  \Acp^{\rm{dir}}(h^+h^-) + \frac{\langle t \rangle}{\tau}\ \Acp^{\rm{ind}}(h^+h^-).
\end{align}
The first term arises from direct \CP\ violation and depends on the decay mode, the second from indirect \CP\ violation and is nearly independent of the decay mode~\cite{Grossman:2006jg}. The average decay time of the sample used in the measurement $\langle t \rangle$ depends on the detector acceptance as a function of decay time and $\tau$ is the $D^0$ lifetime. To measure each individual asymmetry, we determine the number of detected decays of $D^0$ and $\overline{D}^0$ and use the fact that primary charm and anticharm mesons are produced in equal numbers by the  \CP--conserving strong $p\bar{p}$ interactions. 
We require the $D$ candidate to be produced in the decay of an identified $D^{* +}$ or  $D^{* -}$ meson to determine whether the decaying state was initially produced as a $D^0$ or a $\Dbar^0$ meson. Flavor conservation in the  strong-interaction decay of the $D^{*\pm}$ meson, $D^{*+} \to D^0(\to h^+ h^-) \pi^+_s$ and $D^{*-} \to \overline{D}^0(\to h^+ h^-) \pi^-_s$, allows identification of the initial charm flavor through the sign of the charge of the low-momentum $\pi$ meson (soft pion, $\pi_s$). The observed asymmetry, $A(h^+h^-) = N_{\rm{obs}}(D^0\to h^+h^-)-N_{\rm{obs}}(\Dbar^0\to h^+h^-)/$[sum],  is the combination of the contributions from \CP\ violation and from the detection asymmetry between $D^{*+}$ and $D^{*-}$ mesons, due to different reconstruction efficiency for positive and negative soft pions. The combination is linear for small asymmetries, $A(h^+h^-) = \Acp(h^+h^-) + \delta(\pi_s)^{h^+h^-}$. The instrumental asymmetry is due to differences in interaction cross sections with matter between positive and negative low-momentum pions and the geometry of the CDF tracking system (see below).  The combined effect of a few percent cancels in the difference of asymmetries between $K^+K^-$ and $\pi^+\pi^-$ decays,  
\begin{align}
\Delta\Acp =& A(K^+K^-) - A(\pi^+\pi^-)  \nonumber \\* 
		=& \Acp(K^+K^-) - \Acp(\pi^+\pi^-) \nonumber \\*
		 =& \Delta\Acp^{\rm{dir}} +  \Acp^{\rm{ind}}\Delta\langle t \rangle/\tau. \nonumber
		 \end{align}
Kinematic differences between the $K^+K^-$ and $\pi^+\pi^-$ decays result in a fractional 10\% difference in average decay-time of the sample, $\Delta\langle t \rangle/\tau = 0.27 \pm 0.01$, measured through a fit to the decay-time distribution of background-subtracted signal candidates~\cite{Aaltonen:2011se}. Therefore,  most of the indirect \CP-violating asymmetry cancels in the subtraction and $\Delta\Acp$ approximates the difference in direct \CP-violating asymmetries of the two decays.\par
The CDF II detector is a multipurpose magnetic spectrometer surrounded by calorimeters and muon detectors. The detector components relevant for this analysis are briefly outlined below; a more detailed description is in Ref.~\cite{Aaltonen:2011se}. A silicon microstrip vertex detector and a cylindrical drift chamber immersed in a 1.4 T axial magnetic field allow reconstruction of charged-particle trajectories (tracks) in the pseudorapidity range $|\eta| < 1.0$. The vertex detector contains seven concentric layers of single- and double-sided silicon sensors at radii between 1.5 and 22 cm, each providing a measurement with up to 15 (70) $\mu$m resolution in the $\phi$ ($z$) direction~\cite{silicon}. The drift chamber has 96 measurement layers, between 40 and 137 cm in radius, organized into alternating axial and $\pm 2^{\circ}$ stereo superlayers~\cite{COT}. A $35^{\circ}$ tilt angle between the drift chamber cell orientation and the radial direction facilitates track finding, but induces charge-dependent detection asymmetries of up to a few percent for low-momentum charged particles~\cite{Aaltonen:2011se,tesi-angelo}. The component of a charged particle's momentum transverse to the beam ($p_T$) is determined with a resolution of $\sigma_{p_T}/p_T^2 \approx 0.07\%$ ($p_T$ in GeV/$c$), corresponding to a typical mass resolution of 8 MeV/$c^2$ for a two-body charm-meson decay. 
\par The data are collected by a three-level online selection (trigger) system. 
At level 1, tracks are reconstructed in the transverse plane. Two oppositely-charged particles are required, with reconstructed transverse momenta $p_{T1}$, $p_{T2} > 2$ GeV/$c$, the scalar sum $p_{T1} + p_{T2} > 5.5$ GeV/$c$ typically, and an azimuthal opening angle $\Delta \phi < 90^{\circ}$~\cite{XFT}. 
At level 2, tracks are combined with silicon hits and their impact parameter $d$ (transverse distance of closest approach to the beam line) is determined with 45 $\mu$m resolution (including the beam spread) and typically required to be  $0.12 < d < 1.0$ mm~\cite{SVT}. A slightly tighter opening-angle requirement of $2^{\circ}<\Delta\phi<90^{\circ}$  is also applied. Each track pair is then used to form a neutral $D$ candidate, whose flight distance in the transverse plane projected onto the transverse momentum $L_{xy}$ is required to exceed $200$ $\mu$m. At level 3, the selection is confirmed on events that are fully reconstructed by an array of processors.\par 
The offline selection is based on a more accurate determination of the same quantities used in the trigger and is detailed in Ref.~\cite{Aaltonen:2011se,tesi-angelo}.  Here we only describe the improvements specifically aimed at enhancing the sensitivity to $\Delta\Acp$.  By exploiting the highly accurate cancellation of instrumental effects in $\Delta\Acp$, the selection of Ref.~\cite{Aaltonen:2011se} is loosened to significantly increase signal efficiency. Requirements on the minimum number of measurement points for reconstructing tracks are loosened, and the transverse momentum threshold for $D$ decay products is lowered from 2.2 to 2.0 GeV/$c$, nearly doubling the $D^0$ yield. Asymmetries from charm mesons produced in $B$ meson decays (secondary charm) may introduce a common bias in the individual \CP-violating asymmetries, which cancels in the difference of asymmetries.  Secondary charm decays are therefore not excluded from the analysis, providing a 12\% additional  signal. An additional 10\% of events are contributed by secondary trigger selections that were not used in Ref.~\cite{Aaltonen:2011se}. Finally, the additional integrated luminosity contributes approximately 20\% more events. As a result, the expected average resolution on the difference of asymmetries is improved by approximately 50\% with respect to Ref.~\cite{Aaltonen:2011se}.\par
The reconstruction of signal candidates is entirely based on tracking information, with no use of particle identification. Two tracks from oppositely-charged particles, with pion or kaon assignment, are fit to a common decay vertex to form a neutral $D$ candidate. A low-momentum ($p_T>400$ MeV/$c$) charged particle is associated with each $D$ candidate to form the charged $D^*$ candidates.
This allows identification of the charm meson at production as a $D^0$ or a  $\overline{D}^0$ and strongly rejects background, albeit with an 85\% reduction in signal yield.  In the few percent of cases in which multiple candidates per event are reconstructed, one is randomly chosen for further analysis. The resulting sample contains approximately 591~000 $D^0 \to K^+K^-$ candidates, 619~000 $\overline{D}^0 \to K^+K^-$ candidates, 270~000 $D^0 \to \pi^+\pi^-$ candidates, and  279~000 $\overline{D}^0 \to \pi^+\pi^-$ candidates.
Many kinematic distributions are compared for the $K^+K^-$ and $\pi^+\pi^-$ samples.
Small differences are observed in the distributions of the soft pion's impact parameter, transverse momentum, and pseudorapidity. The final candidates are reweighted to make these distributions equal, by using as a weight a smooth function extracted from a fit to the ratio of $K^+K^-$ and $\pi^+\pi^-$ distributions.  The data show that the weight function factorizes into the product of three one-dimensional functions and typically does not exceed 10\%. The reweighting ensures that any kinematics-dependent instrumental asymmetry cancels in the difference of observed asymmetries.\par
The observed asymmetry in each sample is determined from a simultaneous $\chi^2$ fit of the $D^{0}\pi^{+}$ and $\overline{D}^{0} \pi^{-}$ binned mass distributions of candidates restricted to the signal $D$ region, defined as those with mass within 24 MeV/$c^2$ ($3\sigma$) of the known $D^0$ mass. The  $D^0\pi$ mass is calculated using the vector sum of the momenta of the three particles to determine the $D^*$ momentum and the known $D^0$ mass. 
This quantity has the same resolution advantages of  the more customary $M(h^+ h^{-} \pi)- M(h^+ h^{-})$ mass difference, but it is independent of the mass assigned to the $D^0$ decay products. \par The $\pi^+\pi^-$ sample is dominated by the signal of $D^*$--tagged $D^0$ decays, a background of real $D^0$ decays associated with random pions or random combinations of three tracks (combinatorics),  and a 0.93\% contamination of the high-mass tail of the $D^0\to K^-\pi^+$ signal misreconstructed as a $\pi^+\pi^-$ final state. In the $K^+K^-$ sample, an additional background is contributed by mis-reconstructed multi-body charm meson decays, dominated by the $D^0 \to h^-\pi^+ \pi^0$ and the  $D^0 \to h^- \ell^+ \nu_{\ell}$ contributions, where $\ell$ is a muon or an electron.
\begin{figure*}[t]
\centering
\begin{overpic}[width=0.8\textwidth]{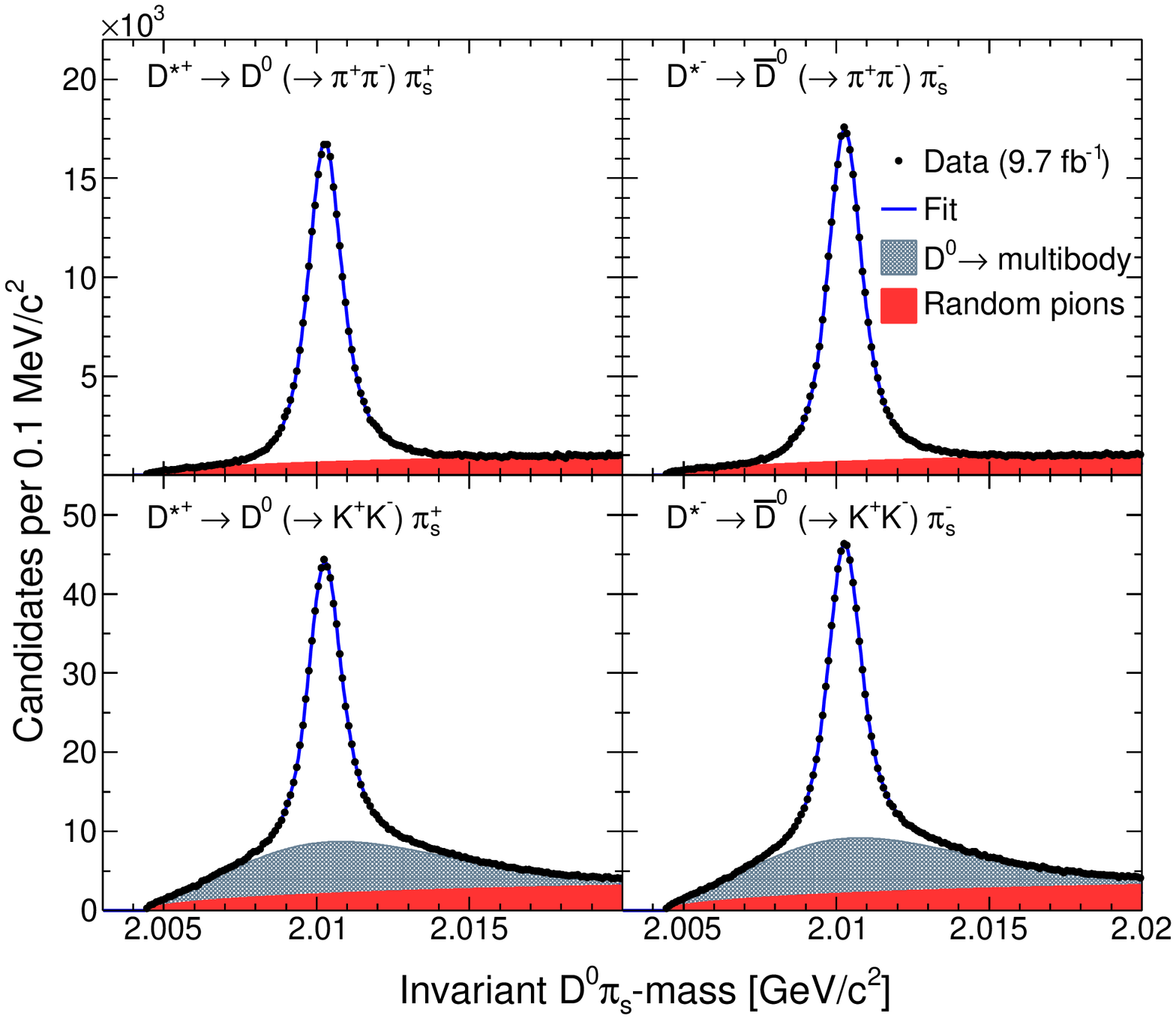}
\put(47,78){(a)}
\put(90,78){(b)}
\put(47,40){(c)}
\put(90,40){(d)}
\end{overpic}
\caption{Distributions of $D^{0}\pi^+$ mass with fit results overlaid for (a) $D^0 \to \pi^+\pi^-$ decays, (b) $\overline{D}^0\to \pi^+\pi^-$ decays, (c) $D^0 \to K^+K^-$ decays, and (d) $\overline{D}^0\to K^+K^-$ decays.}\label{fig:Dstarplots}
\end{figure*}
The functional form of the signal mass shapes is determined from simulation, with parameters tuned in a low-background sample of $12.5\times 10^6$ $D^*$--tagged $D^0\to K^-\pi^+$ decays~\cite{Aaltonen:2011se}. The data indicate a small asymmetry between signal shapes of $D^{*+}$ and $D^{*-}$ decays, which can be attributed to differences in the tracking resolution for positive and negative soft pions. This asymmetry is included in our fit model~\cite{Aaltonen:2011se}. The shape of the combinatorial component is obtained and fixed from data, by forming artificial $D^*$ candidates where each $D^0$ candidate is associated with soft pions of all candidates found in different, randomly chosen events. The functional form of the mis-reconstructed decays is extracted from samples of simulated inclusive charm meson decays, and its parameters are fit together with the desired asymmetries~\cite{Aaltonen:2011se}. The $K\pi$ tail is not included in the $\pi\pi$ fit, but is accounted for in the systematic uncertainties.
In the $\pi^+\pi^-$ sample, the parameters determined by the fit are the asymmetry between $D^{*+}$ and  $D^{*-}$ yields and the relative sizes between the signals and the combinatoric background components. In the $K^+K^-$ sample, the fit also determines the relative sizes and values of the shape parameters of the mis-reconstructed component.
The fit allows for asymmetries between the numbers of combinatorial and mis-reconstructed background events in the  $D^{*+}$ and  $D^{*-}$ samples.
For each final state, we minimize the total $\chi^2$ for the $D^{*+}$ and  $D^{*-}$ samples and obtain the results shown in Fig.~\ref{fig:Dstarplots}. The fits show agreement with data, 
and the observed asymmetries are
\begin{align}
 A(\pi^+\pi^-) = (-1.71 \pm 0.15\stat)\%,   \\
 A(K^+K^-) = (-2.33 \pm 0.14\stat)\%.
 \end{align}
Both asymmetries are dominated by the detector-induced contribution. 
Fits  including extreme variations of the signal and background models yield significantly larger values of reduced $\chi^2$ with minimal variations in the observed asymmetries.
  As a consistency check, we compare the results of the measurement obtained in independent subsamples chosen according to the soft pion's direction in the four quadrants of the tracking volume, different data-taking periods (early and late data), or splitting the present sample in the subsample of events used in Ref.~\cite{Aaltonen:2011se} and the complementary sample, in which we observe $\Delta\Acp = -0.74 \pm 0.27$. The results show a high level of consistency, with reduced $\chi^2$ between observed asymmetry differences of 4.4/3, 0.38/1, and 0.46/1, respectively. \par Most systematic effects cancel in the subtraction of asymmetries.  Residual higher-order instrumental effects that do not cancel are estimated to contribute less than $\pm 0.009\%$ to $\Delta\Acp$, based on simulations
  in which known instrumental asymmetries are varied  as functions of the kinematic variables. The impact of possible residual mis-modeling of the mass shapes used in fits is evaluated by repeating the measurement using extreme variations of the model, as derived from data, and contributes to $\Delta\Acp$ by less than $\pm 0.020\%$. A dominant systematic uncertainty of 0.1\% arises from the possibility that signal and background shapes differ between the  $D^{*+}$ and $D^{*-}$ samples. This effect is assessed by repeating the fit on data using various modifications of the fit shapes in which independent parameters are used for $D^{*+}$ and $D^{*-}$ samples. The effect of the $K\pi$ tail in the $\pi\pi$ signal induces a systematic uncertainty of 0.013\% that is the product of the size of the contamination  (0.93\%) times the 3\% observed asymmetry of the $D^0\to K^-\pi^+$ decay. The impact of the statistical uncertainties associated with the kinematic reweighting  is negligible. \par The final result,  
 \begin{align}
\Delta\Acp = (-0.62 \pm 0.21\stat  \pm 0.10 \syst)\%,
\end{align}
is consistent with and supersedes the previous CDF determination of  $\Delta\Acp = (-0.46 \pm 0.31\stat  \pm 0.12 \syst)\%$~\cite{Aaltonen:2011se}. By adding in quadrature the uncertainties, assumed to be independent and Gaussian-distributed, the difference of asymmetries deviates from zero by $2.7$ standard deviations, strongly indicating the presence of \CP\ violation in the decays of $D^0$ mesons. This result is consistent with the LHCb measurement obtained in $pp$ collisions, $\Delta\Acp = (-0.82 \pm 0.21\stat  \pm 0.11 \syst)\%$~\cite{Aaij:2011in}, with comparable accuracy and less than $1\sigma$ difference in central value. The combined results of the two experiments provide substantial evidence for \CP\ violation in the charm sector with a size larger than most predictions~\cite{SM-predictions}, possibly suggesting the presence of non-SM dynamics. More precise determinations of the individual asymmetries in $D^0 \to \pi^+\pi^-$ and  $D^0 \to K^+ K^-$ decays and extension of the precise experimental exploration to other charm decays  may help in understanding whether the observed effect can be attributed to significant hadronic corrections to the SM weak amplitudes or to new, non-SM sources of \CP\ violation~\cite{theory-interpretations}. 

\begin{acknowledgments}
We thank the Fermilab staff and the technical staffs of the participating institutions for their vital contributions. This work was supported by the U.S. Department of Energy and National Science Foundation; the Italian Istituto Nazionale di Fisica Nucleare; the Ministry of Education, Culture, Sports, Science and Technology of Japan; the Natural Sciences and Engineering Research Council of Canada; the National Science Council of the Republic of China; the Swiss National Science Foundation; the A.P. Sloan Foundation; the Bundesministerium f\"ur Bildung und Forschung, Germany; the Korean World Class University Program, the National Research Foundation of Korea; the Science and Technology Facilities Council and the Royal Society, UK; the Russian Foundation for Basic Research; the Ministerio de Ciencia e Innovaci\'{o}n, and Programa Consolider-Ingenio 2010, Spain; the Slovak R\&D Agency; and the Academy of Finland. 
\end{acknowledgments}


\end{document}

%% file: macro.tex
\newcommand{\Acp}{\ensuremath{A_\textit{CP}}}

\newcommand{\stat}{\ensuremath{\mathrm{~(stat)}}}
\newcommand{\syst}{\ensuremath{\mathrm{~(syst)}}}
\newcommand{\Dbar}{\ensuremath{\overline{D}{}}}

\def \rightdownarrow
 {\kern.3em
 \rule[.5ex]{.15mm}{2ex}
 {\mbox{$\kern-0.1em{\longrightarrow}$}}      
 }

\def\lessim{\mathrel {\vcenter {\baselineskip 0pt \kern 0pt  
\hbox{$<$} \kern 0pt \hbox{$\sim$} }}}

\def\gessim{\mathrel {\vcenter {\baselineskip 0pt \kern 0pt   
\hbox{$>$} \kern 0pt \hbox{$\sim$} }}}

\newcommand{\tev}{\ensuremath{\mathrm{Te\kern -0.1em V}}}
\newcommand{\gev}{\ensuremath{\mathrm{Ge\kern -0.1em V}}}	
\newcommand{\mev}{\ensuremath{\mathrm{Me\kern -0.1em V}}}	
\newcommand{\kev}{\ensuremath{\mathrm{ke\kern -0.1em V}}}	

\newcommand{\CP}{{\it CP}}                                            

\def\beq{\begin{equation}}
\def\eeq{\end{equation}}
\def\bea{\begin{eqnarray}}
\def\eea{\end{eqnarray}}

\def\sss{\scriptscriptstyle}

\def\barp{{\raise.35ex\hbox
{${\sss (}$}}---{\raise.35ex\hbox{${\sss )}$}}}
\def\bdbarp{\hbox{$B_d$\kern-1.4em\raise1.4ex\hbox{\barp}}}
\def\bsbarp{\hbox{$B_s$\kern-1.4em\raise1.4ex\hbox{\barp}}}

\def\roughly#1{\mathrel{\raise.3ex\hbox
{$#1$\kern-.75em\lower1ex\hbox{$\sim$}}}}

\newcommand{\bear}{\begin{array}}
\newcommand{\ear}{\end{array}}
\newcommand{\bet}{\begin{tabular}}
\newcommand{\eet}{\end{tabular}}
\newcommand{\beqn}{\begin{eqnarray}}
\newcommand{\eeqn}{\end{eqnarray}}



%% file: September2011_Authors.tex
\affiliation{Institute of Physics, Academia Sinica, Taipei, Taiwan 11529, Republic of China}
\affiliation{Argonne National Laboratory, Argonne, Illinois 60439, USA}
\affiliation{University of Athens, 157 71 Athens, Greece}
\affiliation{Institut de Fisica d'Altes Energies, ICREA, Universitat Autonoma de Barcelona, E-08193, Bellaterra (Barcelona), Spain}
\affiliation{Baylor University, Waco, Texas 76798, USA}
\affiliation{Istituto Nazionale di Fisica Nucleare Bologna, $^{ee}$University of Bologna, I-40127 Bologna, Italy}
\affiliation{University of California, Davis, Davis, California 95616, USA}
\affiliation{University of California, Los Angeles, Los Angeles, California 90024, USA}
\affiliation{Instituto de Fisica de Cantabria, CSIC-University of Cantabria, 39005 Santander, Spain}
\affiliation{Carnegie Mellon University, Pittsburgh, Pennsylvania 15213, USA}
\affiliation{Enrico Fermi Institute, University of Chicago, Chicago, Illinois 60637, USA}
\affiliation{Comenius University, 842 48 Bratislava, Slovakia; Institute of Experimental Physics, 040 01 Kosice, Slovakia}
\affiliation{Joint Institute for Nuclear Research, RU-141980 Dubna, Russia}
\affiliation{Duke University, Durham, North Carolina 27708, USA}
\affiliation{Fermi National Accelerator Laboratory, Batavia, Illinois 60510, USA}
\affiliation{University of Florida, Gainesville, Florida 32611, USA}
\affiliation{Laboratori Nazionali di Frascati, Istituto Nazionale di Fisica Nucleare, I-00044 Frascati, Italy}
\affiliation{University of Geneva, CH-1211 Geneva 4, Switzerland}
\affiliation{Glasgow University, Glasgow G12 8QQ, United Kingdom}
\affiliation{Harvard University, Cambridge, Massachusetts 02138, USA}
\affiliation{Division of High Energy Physics, Department of Physics, University of Helsinki and Helsinki Institute of Physics, FIN-00014, Helsinki, Finland}
\affiliation{University of Illinois, Urbana, Illinois 61801, USA}
\affiliation{The Johns Hopkins University, Baltimore, Maryland 21218, USA}
\affiliation{Institut f\"{u}r Experimentelle Kernphysik, Karlsruhe Institute of Technology, D-76131 Karlsruhe, Germany}
\affiliation{Center for High Energy Physics: Kyungpook National University, Daegu 702-701, Korea; Seoul National University, Seoul 151-742, Korea; Sungkyunkwan University, Suwon 440-746, Korea; Korea Institute of Science and Technology Information, Daejeon 305-806, Korea; Chonnam National University, Gwangju 500-757, Korea; Chonbuk National University, Jeonju 561-756, Korea}
\affiliation{Ernest Orlando Lawrence Berkeley National Laboratory, Berkeley, California 94720, USA}
\affiliation{University of Liverpool, Liverpool L69 7ZE, United Kingdom}
\affiliation{University College London, London WC1E 6BT, United Kingdom}
\affiliation{Centro de Investigaciones Energeticas Medioambientales y Tecnologicas, E-28040 Madrid, Spain}
\affiliation{Massachusetts Institute of Technology, Cambridge, Massachusetts 02139, USA}
\affiliation{Institute of Particle Physics: McGill University, Montr\'{e}al, Qu\'{e}bec, Canada H3A~2T8; Simon Fraser University, Burnaby, British Columbia, Canada V5A~1S6; University of Toronto, Toronto, Ontario, Canada M5S~1A7; and TRIUMF, Vancouver, British Columbia, Canada V6T~2A3}
\affiliation{University of Michigan, Ann Arbor, Michigan 48109, USA}
\affiliation{Michigan State University, East Lansing, Michigan 48824, USA}
\affiliation{Institution for Theoretical and Experimental Physics, ITEP, Moscow 117259, Russia}
\affiliation{University of New Mexico, Albuquerque, New Mexico 87131, USA}
\affiliation{The Ohio State University, Columbus, Ohio 43210, USA}
\affiliation{Okayama University, Okayama 700-8530, Japan}
\affiliation{Osaka City University, Osaka 588, Japan}
\affiliation{University of Oxford, Oxford OX1 3RH, United Kingdom}
\affiliation{Istituto Nazionale di Fisica Nucleare, Sezione di Padova-Trento, $^{ff}$University of Padova, I-35131 Padova, Italy}
\affiliation{University of Pennsylvania, Philadelphia, Pennsylvania 19104, USA}
\affiliation{Istituto Nazionale di Fisica Nucleare Pisa, $^{gg}$University of Pisa, $^{hh}$University of Siena and $^{ii}$Scuola Normale Superiore, I-56127 Pisa, Italy}
\affiliation{University of Pittsburgh, Pittsburgh, Pennsylvania 15260, USA}
\affiliation{Purdue University, West Lafayette, Indiana 47907, USA}
\affiliation{University of Rochester, Rochester, New York 14627, USA}
\affiliation{The Rockefeller University, New York, New York 10065, USA}
\affiliation{Istituto Nazionale di Fisica Nucleare, Sezione di Roma 1, $^{jj}$Sapienza Universit\`{a} di Roma, I-00185 Roma, Italy}
\affiliation{Rutgers University, Piscataway, New Jersey 08855, USA}
\affiliation{Texas A\&M University, College Station, Texas 77843, USA}
\affiliation{Istituto Nazionale di Fisica Nucleare Trieste/Udine, I-34100 Trieste, $^{kk}$University of Udine, I-33100 Udine, Italy}
\affiliation{University of Tsukuba, Tsukuba, Ibaraki 305, Japan}
\affiliation{Tufts University, Medford, Massachusetts 02155, USA}
\affiliation{University of Virginia, Charlottesville, Virginia 22906, USA}
\affiliation{Waseda University, Tokyo 169, Japan}
\affiliation{Wayne State University, Detroit, Michigan 48201, USA}
\affiliation{University of Wisconsin, Madison, Wisconsin 53706, USA}
\affiliation{Yale University, New Haven, Connecticut 06520, USA}

\author{T.~Aaltonen}
\affiliation{Division of High Energy Physics, Department of Physics, University of Helsinki and Helsinki Institute of Physics, FIN-00014, Helsinki, Finland}
\author{B.~\'{A}lvarez~Gonz\'{a}lez$^z$}
\affiliation{Instituto de Fisica de Cantabria, CSIC-University of Cantabria, 39005 Santander, Spain}
\author{S.~Amerio}
\affiliation{Istituto Nazionale di Fisica Nucleare, Sezione di Padova-Trento, $^{ff}$University of Padova, I-35131 Padova, Italy}
\author{D.~Amidei}
\affiliation{University of Michigan, Ann Arbor, Michigan 48109, USA}
\author{A.~Anastassov$^x$}
\affiliation{Fermi National Accelerator Laboratory, Batavia, Illinois 60510, USA}
\author{A.~Annovi}
\affiliation{Laboratori Nazionali di Frascati, Istituto Nazionale di Fisica Nucleare, I-00044 Frascati, Italy}
\author{J.~Antos}
\affiliation{Comenius University, 842 48 Bratislava, Slovakia; Institute of Experimental Physics, 040 01 Kosice, Slovakia}
\author{G.~Apollinari}
\affiliation{Fermi National Accelerator Laboratory, Batavia, Illinois 60510, USA}
\author{J.A.~Appel}
\affiliation{Fermi National Accelerator Laboratory, Batavia, Illinois 60510, USA}
\author{T.~Arisawa}
\affiliation{Waseda University, Tokyo 169, Japan}
% ...

\collaboration{CDF Collaboration\footnote{With visitors from
$^a$Istituto Nazionale di Fisica Nucleare, Sezione di Cagliari, 09042 Monserrato (Cagliari), Italy,
$^b$University of CA Irvine, Irvine, CA 92697, USA,
$^c$University of CA Santa Barbara, Santa Barbara, CA 93106, USA,
$^d$University of CA Santa Cruz, Santa Cruz, CA 95064, USA,
$^e$Institute of Physics, Academy of Sciences of the Czech Republic, Czech Republic,
$^f$CERN, CH-1211 Geneva, Switzerland,
$^g$Cornell University, Ithaca, NY 14853, USA,
$^h$University of Cyprus, Nicosia CY-1678, Cyprus,
$^i$Office of Science, U.S. Department of Energy, Washington, DC 20585, USA,
$^j$University College Dublin, Dublin 4, Ireland,
$^k$ETH, 8092 Zurich, Switzerland,
$^l$University of Fukui, Fukui City, Fukui Prefecture, Japan 910-0017,
$^m$Universidad Iberoamericana, Mexico D.F., Mexico,
$^n$University of Iowa, Iowa City, IA 52242, USA,
$^o$Kinki University, Higashi-Osaka City, Japan 577-8502,
$^p$Kansas State University, Manhattan, KS 66506, USA,
$^q$Korea University, Seoul, 136-713, Korea,
$^r$University of Manchester, Manchester M13 9PL, United Kingdom,
$^s$Queen Mary, University of London, London, E1 4NS, United Kingdom,
$^t$University of Melbourne, Victoria 3010, Australia,
$^u$Muons, Inc., Batavia, IL 60510, USA,
$^v$Nagasaki Institute of Applied Science, Nagasaki, Japan,
$^w$National Research Nuclear University, Moscow, Russia,
$^x$Northwestern University, Evanston, IL 60208, USA,
$^y$University of Notre Dame, Notre Dame, IN 46556, USA,
$^z$Universidad de Oviedo, E-33007 Oviedo, Spain,
$^{aa}$CNRS-IN2P3, Paris, F-75205 France,
$^{bb}$Texas Tech University, Lubbock, TX 79609, USA,
$^{cc}$Universidad Tecnica Federico Santa Maria, 110v Valparaiso, Chile,
$^{dd}$Yarmouk University, Irbid 211-63, Jordan.
}}
\noaffiliation